 \documentclass[final,5p,times,twocolumn,number]{elsarticle}

\usepackage{amssymb}
\usepackage{amsmath}
\usepackage{lipsum}
\usepackage{hyperref}

\journal{Physics Letters B}

\usepackage[dvipsnames]{xcolor}
\begin{document}

\begin{frontmatter}



\title{QM2-branes and the Swampland Distance Conjecture in nine dimensions}


\author[first,second]{M.P García del Moral}
\affiliation[first]{organization={Departamento de Química, Área de Física, Universidad de la Rioja},
            addressline={
C/ Madre de Dios 53}, 
            city={Logroño},
            postcode={26006}, 
            state={La Rioja},
            country={Spain}}
\affiliation[second]{organization={Instituto de Investigación en Computación Científica (SCRIUR),  Universidad de la Rioja},
            addressline={C/ Madre de Dios 53}, 
            city={Logroño},
            postcode={26006}, 
            state={La Rioja},
            country={Spain}}

\author[third]{C. las Heras}
\affiliation[third]{organization={Instituto de Física y Astronomía, Universidad de Valparaíso},
            addressline={ Av. Gran Bretaña 1111}, 
            city={Valparaíso},
            postcode={2360102}, 
            state={Valparaíso},
            country={Chile}}

\author[fourth,second]{A. Restuccia}
\affiliation[fourth]{organization={Departamento de Física, Universidad de Antofagasta},
            addressline={Av. Universidad de Antofagasta}, 
            city={Antofagasta},
            postcode={02800}, 
            state={Antofagasta},
            country={Chile}}

\begin{abstract}
We discuss the relationship between the Swampland program and QM2-branes, which are supersymmetric M2-branes with discrete spectrum. The worldvolume description of these M2-branes is well-known. The global description is given in terms of twisted torus bundles with monodromy in $\mathrm{SL}(2,\mathbb{Z})$. In particular, we show that the towers of states that ensure the Swampland Distance Conjecture for type II maximal supergravity in nine dimensions are directly related to QM2-branes with trivial monodromy on a background that is toroidally compactified.
\end{abstract}




\end{frontmatter}




\section{Introduction}
\label{sec1}

We analyze the relationship between the Convex Hull Distance Conjecture in type II maximal supergravity in nine dimensions and the M2-brane with fluxes on $M_9\times T^2$. The latter corresponds to supersymmetric M2-branes with a discrete spectrum, whose local and global descriptions are well-known. We show that the towers of states that appear at infinite distances in the moduli space are directly connected to M2-branes with worldvolume fluxes. 

One of the best understood conjectures of the swampland program is the Swampland Distance Conjecture (SDC), proposed in \cite{Ooguri:2006in}. It states that when moving from a point $p_1$ in the moduli space to a point $p_2$ located at infinite distance $d(p_1,p_2)\to \infty$, a tower of states whose mass becomes exponentially light appears.\begin{eqnarray}\label{SDC}m\sim m_0 e^{-\alpha d(p_1,p_2)}\end{eqnarray}where $\alpha$ is a $\mathcal{O}(1)$ factor. In \cite{Etheredge:2022opl}, a sharpened version was proposed, where $\alpha=\frac{1}{\sqrt{d-2}}$, with $d$ the space-time dimensions.

The microscopic nature of the towers is given by the Emergent String Conjecture (ESC) \cite{Lee:2019wij}. It states that every infinite-distance limit in the moduli space of a theory of quantum gravity is either an emergent string limit (featuring a fundamental string with a weakly coupled tower of string oscillation modes) or a decompactification limit (featuring a tower of KK modes). 

The SDC was originally formulated in theories without a scalar potential, but during the last years, many proposals have appeared aiming to explain how the SDC works when there is a scalar potential. See, for example, \cite{Calderon-Infante:2020dhm,Mohseni:2024njl}. Precisely, the Convex Hull Distance Conjecture (CHDC) was originally proposed as a geometric method to understand the SDC in theories with scalar potential \cite{Calderon-Infante:2020dhm}, where not all directions of the moduli space are allowed. The fact that the convex hull associated with the light towers of states in each duality frame lies outside a ball of radius $\alpha_{min}=\frac{1}{\sqrt{d-2}}$ ensures the fulfillment of the Swampland Distance Conjecture. The Convex Hull Distance Conjecture is also a useful tool to study SDC in multi-dimensional moduli spaces. It has been tested in a wide variety of string compactifications \cite{Etheredge:2022opl,Etheredge:2023odp,Calderon-Infante:2023ler,Etheredge:2023usk,Etheredge:2024tok}.

Type II maximal supergravity in nine dimensions corresponds to the simplest string theory compactification for which the Convex Hull Distance Conjecture has been studied. The polytope associated with the convex hull is unique. It has been analyzed from both perspectives, type IIA and type IIB \cite{Etheredge:2022opl,Etheredge:2023odp}. From an M-theory perspective, the three KK vertices of the convex hull correspond to $\frac{1}{2}$ BPS states. One of them is associated with the wrapping of the M2-brane on a torus (which corresponds to $n\neq 0$ and $p,q=0$, with $n$ being the wrapping number and $p,q$ the KK charges), and the other two are related to the KK modes of either 1-cycle of the 2-torus (which corresponds to $n=0$ and $p$ or $q\neq 0$). All of these towers are associated with decompactifying one dimension. Moreover, there are the string oscillators located at the boundary of the sphere with radius $\frac{1}{\sqrt{d-2}}$. In general, at the boundary of the convex hull and at a finite distance in the moduli space, reside the $\frac{1}{4}$ BPS states with both winding and KK modes.

However, in \cite{mpgm23} it is argued that, in the mass operator of the M2-brane on $M_9\times T^2$, and consequently in that of the type IIB ($p,q$)-string on a circle, $p,q\neq 0$ is ensured by a non-trivial wrapping number $n$ of the M2-branes. The issue resides in the map from circles onto circles, which is guaranteed by $n\neq 0$. Such supersymmetric M2-branes have a discrete spectrum \cite{Boulton}. Nevertheless, it is unclear whether supersymmetric M2-branes exhibit good quantum behaviour (i.e.the discreteness of the spectra) in the case of toroidal compactifications with $p,q\neq 0$ and $n=0$.

The action of the supermembrane (M2-brane) formulated on an eleven-dimensional supergravity background was originally proposed in \cite{Bergshoeff,Bergshoeff3}. Other formulations of the supermembrane coupled dynamically to supergravity have been obtained in \cite{Bandos}. The light-cone gauge (LCG) Hamiltonian was found in \cite{deWit2}. The associated $\mbox{SU}(N)$ regularized model has a continuous spectrum \cite{deWit7}. Therefore, it was stated that the M2-brane is unstable. Classically, these instabilities are related to the presence of string configurations, known as spikes, with vanishing energy. This means that M2-branes with different topologies, or any number of membranes, can be related to the same energy \cite{Nicolai}. These spikes are also present in the M2-brane with winding \cite{deWit3,deWit4}. 

Nevertheless, in \cite{Restuccia}, it was found that when the M2-branes have nonvanishing irreducible wrapping on a torus, labeled by the integer $n\ne 0$, the string configurations do not generate any classical instabilities in the theory since they carry energy and they are dynamical excitations. This is because of the presence of a topological restriction on the embedding maps of the theory. The corresponding $\mbox{SU}(N)$ regularized model has a discrete supersymmetric spectrum with finite multiplicity \cite{Boulton}. Since then, different extensions with the same spectral behavior have been found; see, for example, \cite{mpgm6,mpgm25,mpgm26,mpgm27,mpgm28,mpgm29}. We refer to QM2-branes as any of these formulations with discrete spectra. In this work, we will follow the approach of \cite{mpgm6}. We will focus on the M2-brane on $M_9\times T^2$ with worldvolume fluxes, and we will show how the towers that ensure the SDC in type II maximal supergravity in nine dimensions are associated with M2-branes with discrete spectrum.

This work is organized as follows: in section \ref{sec2}, we review the M2-brane with fluxes in $M_9\times T^2$ with a discrete supersymmetric spectrum. In section \ref{sec3}, we use the $T_U$ duality transformation \cite{mpgm5,mpgm7} to identify a new sector of M2-branes with discrete spectra, where the flux condition ensures the presence of the KK term, instead of the wrapping term as in \cite{mpgm6}. In section \ref{sec4}, we show that $1/2$ and $1/4$ BPS towers of the Convex Hull SDC in nine dimensions are associated with sectors of M2-branes with discrete supersymmetric spectrum. Finally, we discuss our results in section \ref{sec5}.

\section{QM2-branes in a nutshell}
\label{sec2}
 In this section, we review the formulation of QM2-branes on a torus, which are supersymmetric M2-branes with discrete spectra. See \cite{mpgm6} for further details.
\subsection{Worldvolume description}
The light cone gauge (LCG) bosonic Hamiltonian for an M2-brane in the presence of a non-vanishing three-form supergravity background was given in \cite{deWit}. Its supersymmetric extension on a Minkowski spacetime, $M_{11}$, was obtained in \cite{mpgm6} where it was also shown its consistency in the presence of a constant $C_{\mu\nu\rho}$ different from zero. 

The corresponding Hamiltonian density is given by 
\begin{eqnarray}\label{HCM2}
\mathcal{H}&=&\frac{1}{2(\widehat{P}_--TC_-)}\left((\widehat{P}_a-TC_a)^2+\frac{T^2}{2}(\epsilon^{uv}\partial_u X^a \partial_v X^b)^2\right) \nonumber \\
&-& T\left(\bar{\theta}\Gamma^-\Gamma_a \left\lbrace X^a,\theta \right\rbrace - C_{+-}- C_+\right), 
\end{eqnarray}
subject to the first and second class constraints
\begin{eqnarray}
\widehat{P}_a\partial_u X^a + \widehat{P}_- \partial_u X^- + \bar{S}\partial_u \theta &\approx& 0, \\
S - (\widehat{P}_--TC_-)\Gamma^- \theta &\approx& 0 ,
\end{eqnarray}
with $T$ being the M2-brane tension and the unique free parameter of the theory, $\widehat{P}_a$ the canonical conjugate to $X^a$ and $S,\bar{S}$ are the conjugate momenta to $\bar{\theta},\theta$ (Majorana spinors in 11D), respectively. The first class constraint is associated with the residual symmetry of the theory, the area preserving diffeomorphisms (APD).

The embedding used in this paper follows the scheme used in seminal papers (see, for example, \cite{deWit2}). In the light-cone gauge, the coordinates split.  The space-time indices $\mu,\nu,\rho=0,\dots,10$ decompose in $\mu=(+,-, a)$, where $a=1,\dots,9$ are the transverse indices to the null light coordinates \cite{deWit2}. The worldvolume indices are labeled by $i=0,1,2$, with $u,v=1,2$ denoting the spatial directions of the worldvolume. We consider an embedding of the M2-brane on the complete 11D space-time. That is,  $X^a(\sigma^1,\sigma^2,\tau)$ are maps from $\Sigma$, the spatial part of the M2-brane worldvolume, which in this case corresponds to a Riemann surface of genus one, to the target space, $X^a: \Sigma \rightarrow M_{11}$.

The pullback of the LCG supergravity three-form components on the M2-brane are written according to \cite{deWit} as
\begin{equation}\label{CaLCG}
\begin{aligned}
& C_a  =  -\epsilon^{uv}\partial_uX^- \partial_vX^b C_{-ab} +\frac{1}{2}\epsilon^{uv}\partial_uX^b \partial_vX^c C_{abc} \, , \\
& C_{\pm}  =  \frac{1}{2}\epsilon^{uv}\partial_uX^a \partial_vX^b C_{\pm ab} \,, \qquad C_{+-}  =  \epsilon^{uv}\partial_uX^- \partial_vX^a C_{+-a} \,,
\end{aligned}
\end{equation}
where $C_{+-a}=0$ is fixed by the gauge invariance of the three-form. By background fixing, the $C_{\pm ab}$, $C_{abc}$ are assumed to be non-trivial constants. In this work, we will fix the background $C_{+ab}=0$ for simplicity. The contribution of the $C_{+ab}$ component has been partially studied in \cite{mpgm23,mpgm6,mpgm24}, and its effect on the convex hull may not be trivial, but this is beyond the scope of the present article and will not be considered any further.

It can be seen that $X^-$ appears explicitly in the Hamiltonian \eqref{HCM2} through the term $C_a$. This dependence introduces non-localities in the formulation, since it is associated with the APD constraints \cite{deWit}. Nevertheless, a canonical transformation of the Hamiltonian can be performed \cite{mpgm6}, and use the residual gauge symmetry generated by the constraints to eliminate the pair ($X^-,P_-^0$) as canonical variables and obtain a formulation solely in terms of the local variables, ($X^a,P_a$) and ($\theta,\bar{S}$).

Now we consider a compactification of the target space on $M_9\times T^2$ with $T^2$ a flat torus characterized by the Teichmuller parameter $\tau\in\mathbb{C}$ with $\mathrm{Im}(\tau)> 0$ and a radius $R\in\mathbb{R}^+$. The embedding maps are now split into the noncompact and compact sectors as follows
\begin{eqnarray*}
X^a(\sigma^1,\sigma^2,\tau)=(X^m(\sigma^1,\sigma^2,\tau),X^r(\sigma^1,\sigma^2,\tau)) , \label{Embedding}
\end{eqnarray*}
with $m=3,\dots,9$ and $r=1,2$. 

We may perform a Hodge decomposition on the closed one-forms $dX^r=dX_h^r + dA^r$, in terms of their harmonic one-forms $dX_h^r$ and their exact parts $dA^r$. $dX_h^r$ may be written in terms of a normalized basis of harmonic one-forms $d\hat{X}^r$ as $dX_h^1+idX_h^2 = 2\pi R(l_r+m_r\tau)d\hat{X}^r$. The wrapping condition on the compact sector is given by
\begin{equation}
\label{ec2notas}
\oint_{\mathcal{C}_r} d \left(X^1 + iX^2 \right)= 2 \pi R \left(l_r + m_r \tau \right) \, \in  \mathcal{L} \,,
\end{equation}
where $\mathcal{C}_r$ denotes the homology basis on $\Sigma$, $\mathcal{L}$ is a lattice on the complex plane ($\mathbb{C}$) such that $T^2=\mathbb{C}/\mathcal{L}$ and $l_r,m_r$  denote the wrapping numbers that define the wrapping matrix
 \begin{eqnarray}
     \mathbb{W} = \begin{pmatrix}
        l_1 & l_2 \\
          m_1 & m_2
          \end{pmatrix} .    
 \label{windingmatrix}
 \end{eqnarray}
So far, the Hamiltonian of the M2-brane on a torus can be written as
\begin{eqnarray}
H^{T^2}&=&\frac{1}{2P^0_-}\int_\Sigma d^2\sigma \sqrt{w}\left[\Big(\frac{P_m}{\sqrt{w}}\Big)^2+\Big(\frac{P_r}{\sqrt{w}}\Big)^2 \right. \nonumber \\
&+&\left. \frac{T^2}{2}\left(\left\{X^m,X^n\right\}^2 + 2\left\{X^m,X^r\right\}^2 + \left\{X^r,X^s\right\}^2\right)\right]\nonumber \\
&-& \frac{T}{2P^0_-}\int_\Sigma d^2\sigma \sqrt{w} (\bar{\theta}\Gamma_-\Gamma_r\left\{X^r,\theta\right\}-T\bar{\theta}\Gamma_-\Gamma_m\left\{X^m,\theta\right\}), \nonumber \\
\label{HamiltonianM2}
\end{eqnarray}
where $\displaystyle \left\lbrace \bullet, \bullet \right\rbrace = \frac{\epsilon^{uv}}{\sqrt{w}}\partial_u \bullet \partial_v \bullet$ and $P_-=P_-^0\sqrt{w}$. 

Now, let us make some comments on this Hamiltonian. The harmonic maps on the compact sector may degenerate. Moreover, the string spikes identified of the M2-brane in $M_{11}$ and in $M_{10}\times S^1$, are also present in this torus compactification. Consequently, it has a
continuous spectrum \cite{deWit7,deWit3,deWit4}. These two problems have been studied during the last decades (see, for example, \cite{Boulton,Restuccia, mpgm6,Restuccia3}) obtaining a condition for the discreteness of the mass operator of the supersymmetric M2-brane -which will be explained in the following paragraphs- by ensuring the non-vanishing of the determinant of the wrapping matrix, $\det( \mathbb{W})=n$ with $n\in \mathbb{Z}-\{0\}$. This is known as the \textit{irreducible} wrapping condition.

Once the dependence on $X^-$ has been eliminated, a quantization condition on $C_{-}$ can be imposed. This condition corresponds to a 2-form flux condition on the target space 2-torus, whose pull-back through $X_h^r$, with $r=1,2$, generates a 2-form flux condition on the M2-brane worldvolume as follows \cite{mpgm10}
 \begin{eqnarray}\label{fluxpullback}
 \int_{T^2}C_{-}&=&\frac{1}{2} \int_{T^2}C_{- rs} d\widetilde{X}^r\wedge d\widetilde{X}^s \nonumber \\ 
 &=&  c_{-}\int_\Sigma \widehat{F} \nonumber \\
 &=& k_{-},
 \end{eqnarray}
 where $C_{- rs}=c_{-}\epsilon_{rs}$ with $c_{-}\in \mathbb{Z}-\{0\}$, $\widetilde{X}^r$ are local coordinates on $T^2$, $k_{-}=nc_{-}$. It can be seen that the closed 2-form  $\widehat{F}$ is defined on $\Sigma$ such that it describes a worldvolume flux condition
 \begin{equation}\label{central charge}
   \int_{\Sigma}\widehat{F} = \frac{1}{2}\int_{\Sigma}\epsilon_{rs}d\widehat{X}^r \wedge d\widehat{X}^s =n,
  \end{equation}
  where the integer $n\ne 0$ characterizes the irreducibility of the wrapping. Consequently, $C_{-}$ is a closed two-form defined on the target space torus. Indeed, the flux condition on $T^2$ implies a flux condition on $\Sigma$ which is known as the `central charge condition'. The irreducible wrapping condition ensures that the harmonic modes are nontrivial and independent.

The Hamiltonian of the M2-brane with $C_-$ fluxes becomes
\begin{eqnarray}
H^{C_-}&=&\frac{1}{2P^0_-}\int_\Sigma d^2\sigma \sqrt{w}\left[\Big(\frac{P_m}{\sqrt{w}}\Big)^2+\Big(\frac{P_r}{\sqrt{w}}\Big)^2   \right. \nonumber \\
&+&\left. \frac{T^2}{2}\left(\left\{X^m,X^n\right\}^2 +  2(\mathcal{D}_rX^m)^2+(\mathcal{F}_{rs})^2+ (\widehat{F}_{rs})^2\right)\right] \nonumber \\ &-& \frac{T}{2P^0_-}\int_\Sigma d^2\sigma \sqrt{w} (\bar{\theta}\Gamma_-\Gamma_r\mathcal{D}_r\theta-T\bar{\theta}\Gamma_-\Gamma_m\left\{X^m,\theta\right\}). \nonumber \\ \label{HamiltonianM2NT}
\end{eqnarray}

 Interestingly, it is completely equivalent to consider a supermembrane on $M_9\times T^2$ with a central charge condition associated with an irreducible wrapping \cite{Restuccia}, and an M2-brane formulated in the same space-time with a quantized $C_-$ form. Their associated Hamiltonians are exactly the same, $\mathcal{H}^{CC}=\mathcal{H}^{C_-}$.
The degrees of freedom of the theory are single-valued and correspond to $X^m,A^r,\theta$. The embedding considered here is completely general. Therefore, although the M2-branes wrap the 2-torus, they are also extended membranes in the non-compact space rather than just free particles. They contain ampng other terms, a non-linear quartic bosonic potential and its fermionic counterpart.

The symplectic covariant derivative in the Hamiltonian $H^{C_-},$ is defined as \cite{Ovalle1}
\begin{eqnarray}
   \mathcal{D}_rX^m &=&D_rX^m+\left\{ A_r,X^m\right\}, \label{symp-cov-der}
\end{eqnarray}
with $D_r$ is a covariant derivative defined as in \cite{mpgm2,mpgm7} and satisfies 
\begin{eqnarray}
    (D_1+iD_2) \, \bullet  = 2\pi R (l_r+m_r\tau)\left\lbrace \widehat{X}^r,\, \bullet \right\rbrace . \nonumber 
\end{eqnarray}
 The worldvolume flux contribution  (\ref{fluxpullback}) in the Hamiltonian \eqref{HamiltonianM2NT} is associated with 
 $\widehat{F}_{rs}$. The symplectic gauge curvature,
\begin{eqnarray}
  \mathcal{F}_{rs}&=& D_rA_s-D_sA_r+\left\{ A_r,A_s\right\}, \label{Fsymp}
\end{eqnarray}
is associated with the one-form connection, $A_r dX^r$, where the $A^r$ are dynamical degrees of freedom related to the exact sector of the map on $T^2$.

This Hamiltonian is subject to the local and global constraints associated with the APD, which are isomorphic to the symplectomorphisms on a two-dimensional manifold, 
\begin{eqnarray}
\left\{ \frac{P_m}{\sqrt{w}} , X^m\right\} + \mathcal{D}_r\left( \frac{P_r}{\sqrt{w}}\right)+\left\lbrace \frac{\bar{S}}{\sqrt{w}},\theta \right\rbrace  &\approx& 0, \label{LocalAPD}\\
 \oint_{C_S}\left[\frac{P_m dX^m}{\sqrt{w}} + \frac{P_r (dX_h^r+dA^r)}{\sqrt{w}} + \frac{\bar{S} d\theta}{\sqrt{w}}\right] &\approx& 0. \label{GlobalAPD}
\end{eqnarray}
They appear as a residual symmetry on the theory after imposing the LCG in the covariant formulation. We have shown that M2-branes with $C_{-}$ fluxes are invariant under the full group of symplectomorphisms, which considers the sectors connected and not connected to the identity. Furthermore, symplectomorphisms on $T^2$ are in one-to-one correspondence to symplectomorphisms on $\Sigma$  \cite{mpgm10}. 
Classically, this Hamiltonian does not contain string-like spikes at zero cost energy that may produce instabilities as shown in \cite{mpgm}. At the quantum level,  $\mathrm{SU}(N)$ regularized theory has a purely discrete spectrum since it satisfies the sufficiency criteria for discreteness found in \cite{Boulton}.  The theory preserves half of the supersymmetry in a minimal configuration involving either Kaluza-Klein or wrapping charges. Thus, it preserves a quarter of the supersymmetry in a more general state involving both types of charges within the interior of the moduli space, as discussed in \cite{mpgm6}.

According to \cite{mpgm19}, the global formulation of these M2-branes is realized in terms of twisted torus bundles with monodromy in $\mathrm{SL}(2,\mathbb{Z})$. 
In nine dimensions, the symmetry group of type IIB gauge supergravities is reproduced by the equivalence classes of bundles related to the nontrivial monodromies.  In this paper, we will only consider the case of trivial monodromy.

\subsection{Mass operator of the supermembrane with $C_{-}$ fluxes}
First, let us compute the winding and KK terms of the mass operator \cite{mpgm23}. The embedding map to the compact sector is defined as  
\begin{eqnarray}\label{generalmap}
 dX=(2\pi R)(l_s+m_s\tau)d\hat{X}^s+dA, 
 \end{eqnarray}
 where $dA=dA^1+idA^2$ is a dynamical exact one-form. However, this expression can be rewritten using the independent and arbitrary $\mathrm{SL}(2,Z)$ symmetries on $T^2$ and $\Sigma$ as $dX=2\pi R (n d\hat{X}^{1}+ \tau d\hat{X}^{2})+dA$.

 The Hamiltonian contribution to the wrapping on the M2-brane is expressed in terms of the harmonic one-form as follows, 
 \begin{eqnarray}
    \frac{T^2}{P_-^0} \int d^2\sigma \left[\frac{1}{4}\sqrt{w}\left\lbrace X_h^r,X_h^s \right\rbrace^2 \right] = \frac{1}{2P_-^0}(TnA_{T^2})^2,
 \end{eqnarray}
  with $n=\det(\mathbb{W})$. Therefore, the wrapping term on the mass operator of the M2-brane is directly given by the $C_-$ fluxes \eqref{fluxpullback} -equivalently, the central charge contribution- 
  \begin{eqnarray} \label{massoperatorwinding}
    M_{C_-}^2 = (TnA_{T^2})^2 + \dots.
 \end{eqnarray}
 As the irreducible wrapping condition guarantees that $n\neq 0$, the winding term is directly related to these sectors. 
 
 To reproduce the KK term on the mass operator, we recall the definition of the zero modes of the momentum in the compact sector as 
 \begin{eqnarray}
P^0_{r}=\int_\Sigma p_r d\sigma^1\wedge d\sigma^2,
\end{eqnarray}
with $r=1,2$. They can be expressed in terms of the Hodge dual of two well-defined associated 2-forms $(F)^r$ on $\Sigma$. In fact, fixing $r$, it can be seen that
\begin{eqnarray}
Rp&=& \sqrt{w}\left(\star F\right),
\end{eqnarray}
with $\displaystyle \star F = \frac{\epsilon^{uv}F_{uv}}{2\sqrt{w}}$.

Consequently,
\begin{eqnarray}
RP^0 &=& \int_\Sigma F,
\end{eqnarray}
and then, the following quantization conditions are imposed for each value of $r$
\begin{eqnarray}\label{FluxKK}
\int_\Sigma F_r = \widehat{m}_r \in \mathbb{Z}.
\end{eqnarray}
To guarantee that the maps from the base manifold to the compact target sector are from circles onto circles, we must consider the left-hand member of
\begin{eqnarray}\label{circletocirlce}
\frac{1}{2\pi R}\oint_{C_S}\mathbb{M}^{-1}\begin{pmatrix}
    dX^1 \\
    dX^2
\end{pmatrix} = \oint _{C_S}\mathbb{W}\begin{pmatrix}
    d\widehat{X}^1 \\
    d\widehat{X}^2
\end{pmatrix},
\end{eqnarray}
with
\begin{eqnarray}
\mathbb{M} &=& \begin{pmatrix}
    1 & \mathrm{Re}(\tau) \\
     0 & \mathrm{Im}(\tau)
\end{pmatrix}\label{matrixM}, \\
\mathbb{W} &=& \begin{pmatrix}
    n & 0 \\
     0 & 1
\end{pmatrix}\label{matrixM}, 
\end{eqnarray}
such that $\det (\mathbb{W})=n\neq 0$. Now, by using the corresponding conjugate momenta, 
\begin{eqnarray}\label{Conju_Momenta}
RP^0_s\mathbb{M}^s_r=R\int_\Sigma p_s\mathbb{M}^s_rd\sigma^1\wedge d\sigma^2 = \widehat{m}_r.
\end{eqnarray}
Consequently, the KK modes are given by
$
\displaystyle P^0_r = (\mathbb{M}^{-1})^s_{r}\frac{\widehat{m}_s}{R}.
$
Hence,
\begin{eqnarray}
P^0_1 = \frac{\widehat{m}_1}{R},\label{P01} \quad P^0_2 = \frac{\widehat{m}_2-\widehat{m}_1\mbox{Re}(\tau)}{R\mbox{Im}(\tau)}\, .\label{P02}
\end{eqnarray}
 The KK contribution to the mass operator is given by \cite{mpgm19}
\begin{eqnarray}\label{KKterma}
2P^0_- \left( \frac{1}{2P^0_-}P^0_rP^{0r}\right) =  m^2\frac{\vert q\tau-p \vert^2 }{(R\mbox{Im}(\tau))^2} =  \frac{m^2}{Y^2} ,
\end{eqnarray}
with $p,q$ relatively primes and 
\begin{eqnarray}\label{CoordY}
Y=\frac{R\mathrm{Im}\tau}{\left\vert q\tau-p \right\vert}.    
\end{eqnarray}

We have also used that $\widehat{m}_1=mp$ and $\widehat{m}_2=mq$, with $m\in\mathbb{Z}$. 


A not-so-obvious aspect is that implicitly, the arbitrariness in the allowed values of $p,q$ is directly related to a well-defined compactification on a 2-torus, i.e., it is associated with the irreducible wrapping condition present in well-behaved sectors of M2-branes. Indeed, when the wrapping of the M2-brane on the compact sector is reducible, $det(\mathbb{W})=n=0$, the map from $\Sigma$ to $T^2$ becomes degenerate and consequently, there is no map from circles to circles, i.e., $p,q\neq 0$ is not possible \cite{mpgm23}.

Let us emphasize that $\widehat{m}_r$ in \eqref{FluxKK} can be zero and is still consistent with $n\neq 0$, leading to an M2-brane with wrapping term and no KK contribution. This comment will be extended in the following section.

 Finally, the M2-brane with $C_-$ fluxes mass operator corresponds to \cite{mpgm23} 
\begin{eqnarray}\label{MassOp_MonTriv}
    \mathcal{M}_{C_-}^2 &=& (TnA_{T^2})^2 +  \frac{m^2}{Y^2}+2\widehat{P}_-^0H'^{C_-},
 \end{eqnarray}
 with the Hamiltonian associated with the nonzero modes is given by
\begin{eqnarray}
H'^{C_-}&=&\frac{1}{2P^0_-}\int_\Sigma d^2\sigma \sqrt{w}\left[\Big(\frac{P'_m}{\sqrt{w}}\Big)^2+\Big(\frac{P'_r}{\sqrt{w}}\Big)^2 \right. \nonumber \\
&+&\left. \frac{T^2}{2}\left(\left\{X^m,X^n\right\}^2 + 2(\mathcal{D}_rX^m)^2(\mathcal{F}_{rs})^2\right)\right] \nonumber \\
&-& \frac{T}{2P^0_-}\int_\Sigma d^2\sigma \sqrt{w} (\bar{\theta}\Gamma_-\Gamma_r\mathcal{D}_r\theta-T\bar{\theta}\Gamma_-\Gamma_m\left\{X^m,\theta\right\}),  \nonumber \\ \label{Hamiltonian_prime_Cmenos}
\end{eqnarray}
where the prime on the fields in the Hamiltonian indicates that the zero modes -associated with the center of mass- have been excluded.

The wrapping and KK contribution on the Hamiltonian were obtained in \cite{Schwarz6}. They are directly related to the M2-brane with a central charge condition associated with the irreducibility of the wrapping or with the presence of $C_-$ fluxes, on $M_9\times T^2$. The worldvolume flux condition ensures the appearance of both terms. Moreover, it allows  to compute the membrane excitations from the worldvolume theory of the M2-brane.
\subsection{Associated states in type IIB string theory}

The double dimensional reduction of the mass operator \eqref{MassOp_MonTriv}, leads to the mass operator of a $(p,q)$-string on a circle
\begin{eqnarray}
M^2_{(p,q)}= \left(\frac{n}{R_B}\right)^2+\left( 2\pi R_B m T_{(p,q)}\right)^2
+ 4\pi T_{(p,q)}(N_L+N_R), \label{massoppq} 
\end{eqnarray}
where 
\begin{eqnarray}
    T_{(p,q)}&=&\frac{\left\vert q\lambda_0-p\right\vert}{(\mbox{Im}\lambda_0)^{1/2}}T_c, \\
    R_B^2 &=& \left(TA_{T^2}T_c\right)^{-1},
\end{eqnarray}
and $\tau=\lambda_0$, being $\lambda_0$ the type IIB axio-dilaton.

The winding and KK terms were found in \cite{Schwarz}. In that reference, the oscillators term was not computed, but it was mentioned that they were related to membrane excitations. Nevertheless, in \cite{mpgm23} it is shown that the winding and KK terms are directly related to the M2-brane with worldvolume fluxes on $M_9\times T^2$, and the oscillators term was computed from the nonzero modes of the M2-brane Hamiltonian.

Because of T-duality, let us note that the integer $n$, which is the wrapping number in M2-brane theory, corresponds to the KK number in type IIB string theory. Equivalently, $m$ associated with the KK charges from M-theory is the winding number from the type IIB perspective.



\section{(Not so) Novel QM2-branes}
\label{sec3}

In this section, we describe a new sector of supersymmetric M2-branes with a discrete spectrum. This sector is characterized by a flux condition ensuring the appearance of the KK term, instead of the winding as \eqref{fluxpullback}. It is associated with the $T_U$-dual of the M2-brane with central charge but no KK term \cite{mpgm7}.
\subsection{$T_U$ duality transformation}


The action of U-duality in QM2-branes is well-known. In particular. The $\mathrm{SL}(2,\mathbb{Z})$ transformation corresponds to the symplectomorphisms of the 2-torus of the target space and the monodromy of the bundle. The $T_U$-duality transformation acts on the mass operator and on the bundle description.

From \cite{mpgm5,mpgm7}, we know that the $T_U$-duality transformation on quantum M2-branes is given by a matrix
\begin{equation}
    \mathcal{T}= \begin{pmatrix}
        \alpha & \beta \\
        \gamma & \alpha
    \end{pmatrix} \in \mbox{SL}(2,\mathbb{Z}).
\end{equation}
It acts in the mass operator of the M2-brane with $C_-$ fluxes, given by \eqref{HamiltonianM2NT}, as
\begin{eqnarray}
    \mathcal{Z} \to \widetilde{\mathcal{Z}}=\frac{1}{\mathcal{Z}} , \quad  \tau\to\widetilde{\tau}=\frac{\alpha\tau+\beta}{\gamma\tau+\alpha}\\
\mathbb{W}\to\widetilde{\mathbb{W}}=\mathcal{T}\mathbb{Q} , \quad  \mathbb{Q}\to\widetilde{\mathbb{Q}}=\mathcal{T}^{-1}\mathbb{W}  
\end{eqnarray}
with $\mathcal{Z}=\left(TA_{T^2}Y\right)^{1/3}$ an adimensional variable proportional to $R$, $A_{T^2}=(2\pi R)^2\mathrm{Im}\tau$ and $Y$ given by \eqref{CoordY}. 

It can be checked that $\widetilde{\mathcal{Z}}\mathcal{Z}=1$, implies
\begin{eqnarray}
    R \to \widetilde{R}=\frac{\left\vert \gamma\tau + \alpha\right\vert \left\vert q\tau - p\right\vert^{2/3}}{T^{2/3}(2\pi)^{4/3}\mbox{Im}(\tau)^{4/3}R}, 
\end{eqnarray}
and such transformation reproduces the known $R\to\frac{\alpha'}{R}$ behaviour under double dimensional reduction \cite{mpgm5}.

 The mass operator of the dual theory is given by
\begin{eqnarray}
\widetilde{\mathcal{M}}^2
&=& \left( Tn\widetilde{A}_{T^2} \right)^2 + \left( \frac{m}{\widetilde{Y}} \right)^2 +2\widehat{P}^2_0 \widetilde{H'}\nonumber \\
&=& \frac{1}{\mathcal{Z}^2}\left(\frac{n}{Y}\right)^2 + \frac{1}{\mathcal{Z}^2}\left(TmA_{T^2}\right)^2+ 2\widehat{P}^2_0 \frac{1}{\mathcal{Z}^8}H'.
\end{eqnarray}

The $T_U$-duality transformation on the M2-brane with $C_-$ fluxes maps winding to KK, and viceversa, as expected. 

It can be seen that the mass operator is invariant when $\mathcal{Z}=1$ and the values of the integers $n$, $m$ associated with the KK and central charge terms are interchanged 
\begin{eqnarray}
\widetilde{\mathcal{M}}^2=\mathcal{M}^2.
\end{eqnarray}

In this scenario, the tension of the QM2-brane is fixed in terms of the moduli and KK charges
\begin{eqnarray}
    T=\frac{1}{(2\pi)^2}\frac{\left\vert q\tau-p \right\vert}{R^3\left(\mathrm{Im}\tau\right)^2}.
\end{eqnarray}

In the global description, $T_U$-duality transforms a bundle into a dual one by interchanging the topological invariants, while the geometry is mapped to the dual moduli. 

\subsection{M2-brane with fluxes, discrete spectrum and vanishing wrapping number}

In general, the M2-brane with fluxes in $M_9\times T^2$ is characterized by $n\in\mathbb{Z}-\left\lbrace 0 \right\rbrace$ \eqref{fluxpullback}. This integer corresponds to the wrapping number in \eqref{MassOp_MonTriv}. Nevertheless, it can be seen that $m_1,m_2\in\mathbb{Z}$ are associated with the map from circles onto circles \eqref{circletocirlce}. Consequently, when $m_1=m_2=0$, we have that the mass operator is given by
\begin{eqnarray}
\mathcal{M}^2 &=& \left(TnA_{T^2}\right)^2 + 2\widehat{P}^2_0 H',
\end{eqnarray}
and the $T_U$ dual mass operator corresponds to
\begin{eqnarray}\label{massopdual}
\widetilde{\mathcal{M}}^2 &=&  \frac{1}{\mathcal{Z}^2}\left(\frac{n}{Y}\right)^2 + 2\widehat{P}^2_0 \frac{1}{\mathcal{Z}^8}H'.
\end{eqnarray}
 Indeed, the action of the $T_U$ duality transformation on the flux condition \eqref{fluxpullback} is the following
 \begin{eqnarray}
     \frac{1}{2\pi}\int F= nA_{T^2} \to \frac{1}{2\pi}\int F= n\widetilde{A}_{T^2} = \frac{1}{\mathcal{Z}}\frac{n}{Y},  \label{dualflux}
 \end{eqnarray}
 with $F=\frac{1}{2}\epsilon^{rs}   dX^r\wedge dX^s$ invariant.

 The mass operator \eqref{massopdual} corresponds to an M2-brane -with worldvolume fluxes, a KK contribution, and vanishing wrapping-, which is $T_U$-dual to the M2-brane with central charge and no KK modes. Consequently, this sector of supersymmetric M2-branes has a discrete spectrum. 
 
While in \cite{Restuccia,mpgm6} the worldvolume fluxes ensure that the wrapping number is nonvanishing, in the dual theory, it ensures that the KK term is nonvanishing, as it can be seen from \eqref{dualflux}.

It is worth saying that this sector is implicitly considered in \cite{mpgm5,mpgm7}, where the $T_U$-duality transformation on the M2-branes is introduced. The sector of QM2-branes introduced in this subsection corresponds to a particular scenario.

\subsection{Associated states in type IIB string theory}

From \eqref{massoppq}, we note that the M2-brane with central charge and no KK contribution reproduces the bound states of type IIB ($p,q$)-strings on a circle with vanishing winding and nonvanishing KK contribution
\begin{eqnarray}
M^2_{(p,q)}= \left(\frac{n}{R_B}\right)^2
+ 4\pi T_{(p,q)}(N_L+N_R). \label{massoppq2} 
\end{eqnarray}
Double-dimensional reduction of the mass operator \eqref{massopdual} of the $T_U$ dual theory leads to 
\begin{eqnarray}
    M^2_{(p,q)}= \left( 2\pi R_B n T_{(p,q)}\right)^2
+ 4\pi T_{(p,q)}(N_L+N_R), \label{massoppq3}
\end{eqnarray}
for $\mathcal{Z}=1$. It corresponds to
bound states of $(p,q)$-strings on a circle with winding, but with vanishing KK numbers. Let us note that, in both cases, the tower and the oscillator term are directly related to $n\in\mathbb{Z}-\left\lbrace 0 \right\rbrace$ ensured by the worldvolume flux condition \eqref{fluxpullback}.

These mass operators are associated to states $\frac{1}{2}$ BPS, while the more general states with winding and KK contribution, given by \eqref{massoppq}, correspond to states $\frac{1}{4}$ BPS. Both states, $\frac{1}{2}$ and $\frac{1}{4}$ BPS are directly related with supersymmetric M2-branes with discrete spectrum.

\section{QM2-branes and the Swampland Distance Conjecture}
\label{sec4}
In this section, we discuss the origin in M-theory of the towers ensuring the SDC conjecture in type II maximal supergravity in nine dimensions. We found that all the towers are associated with supersymmetric M2-branes with discrete spectrum, whose worldvolume and bundle description are well-known.

\subsection{The convex hull distance conjecture}

The Swampland Distance Conjecture rest at the foundations of the Swampland program. It has been tested in a great variety of backgrounds (see, for example, \cite{Etheredge:2023odp,Grimm:2018ohb,Font:2019cxq,Corvilain:2018lgw,Ashmore:2021qdf,Erkinger:2019umg,Joshi:2019nzi}). A geometrical proposal to understand the SDC, even in theories with scalar potential, is the Convex Hull Distance Conjecture. It has been explored in many different compactifications, giving a geometric picture associated with the towers that ensure the SDC \cite{Etheredge:2022opl,Etheredge:2023odp,Etheredge:2023usk,Etheredge:2024tok}.

Let us consider a theory in $d$ dimensions with an action given by
\begin{eqnarray}
S=M_{\mbox{Pl};\mbox{d}}^{d-2}\int d^dx\sqrt{-h}\left(\frac{R}{2}-\frac{1}{2}G_{ij}\partial_\mu\phi^i\partial^\mu\phi^j+\dots\right),
\end{eqnarray}
where $\phi^i$ corresponds with a series of massless scalar fields weakly coupled to gravity, and $G_{ij}$ is the field space metric. The geodesic field distance is given by
\begin{eqnarray}
d(\phi_0,\phi)=\int_{\phi_0}^\phi\sqrt{G_{ij}d\phi^id\phi^j}.
\end{eqnarray}
Let us define a scalar charge-to-mass vector of particle of mass $M$ \cite{Calderon-Infante:2020dhm,Etheredge:2023odp} 
\begin{eqnarray}
    z_{i} &=& -\partial_i \log M,
\end{eqnarray}
where $\partial_i=\frac{\partial}{\partial\phi^i}$ and $i$ runs from zero to the number of independent moduli. The dot product is given by $z^2=G^{ij}z_iz_j$. The projection of $z_i$ along a geodesic trajectory approaching the asymptotic region of the moduli space gives the exponential rate at which its  mass decreases
\begin{eqnarray}
    \alpha=\Vec{z}\cdot \Vec{\eta},
\end{eqnarray}
with $\eta^a=e_i^a\frac{\partial_\lambda \phi^i(\lambda)}{\vert \partial_\lambda \Vec{\phi} \vert}$ is the normalized tangent vector to the asymptotic trajectory. 

The Convex Hull Distance Conjecture (CHDC) states that in any given asymptotic region of a quantum gravity theory, the outside boundary of the convex hull generated by the $\vec{\zeta}$-vectors of all light towers must remain outside the ball of radius $\alpha_{min}$ in the range of directions defining the asymptotic region. 
    
\subsection{The CHDC in type II maximal supergravity}

\begin{figure}
    \centering
\includegraphics[width=0.7\linewidth]{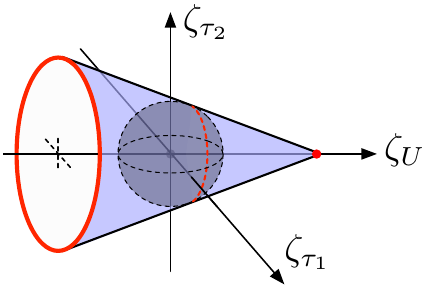}
    \caption{Convex Hull of type II maximal supergravity from the type IIA perspective taken from \cite{Etheredge:2022opl}}
    \label{9dCHIIA}
\end{figure}

\begin{figure}
    \centering
\includegraphics[width=0.7\linewidth]{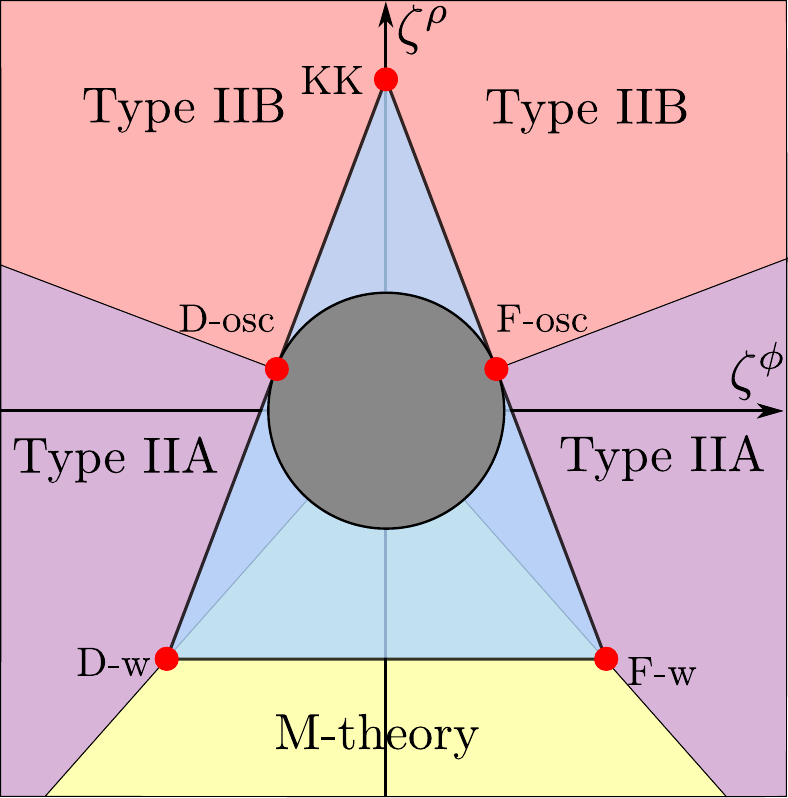}
    \caption{Convex Hull of type II maximal supergravity from the type IIB perspective taken from \cite{Etheredge:2023odp}}
    \label{9dCHIIB}
\end{figure}

\begin{figure}
    \centering
\includegraphics[width=0.7\linewidth]{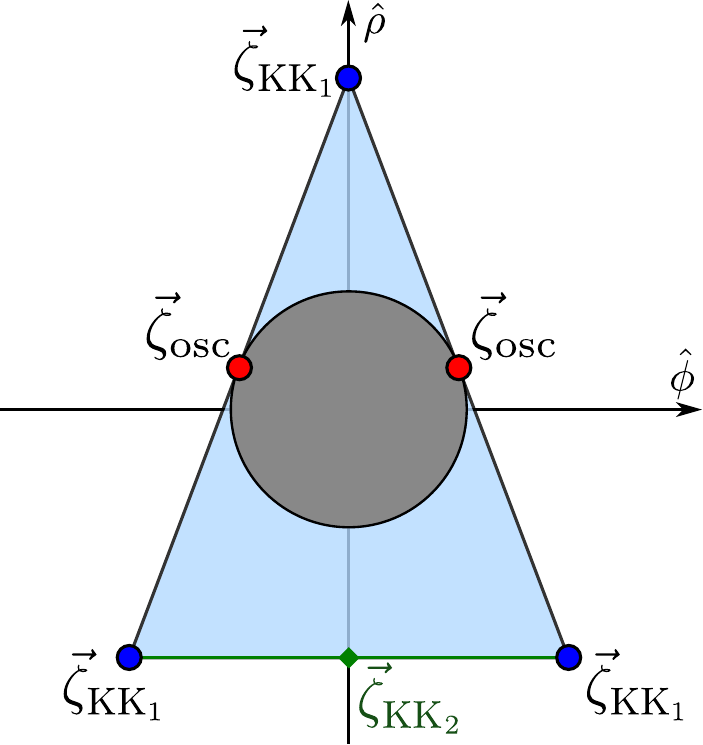}
    \caption{Convex Hull of type II maximal supergravity from the type M-theory perspective taken from \cite{Etheredge:2024tok}}
    \label{9dM}
\end{figure}

One of the simplest compactifications where the Convex Hull was studied is on type II maximal supergravity in nine dimensions. In this case, the polytope associated with the convex hull is unique. It has been studied from the type IIA, type IIB and M-theory perspective \cite{Etheredge:2022opl,Etheredge:2023odp,Etheredge:2024tok}. See figures: \ref{9dCHIIA}, \ref{9dCHIIB}, and \ref{9dM} taken from taken from \cite{Etheredge:2022opl}, \cite{Etheredge:2023odp}, and \cite{Etheredge:2024tok}, respectively. 

The towers appearing at infinite distances of the moduli space define a cone as the convex hull when $\mathrm{Re}(\tau)\neq 0$ in the space of the scalar charge-to-mass vectors. Inside the convex hull, there is a ball of radius $\alpha_{min}=\frac{1}{\sqrt{d-2}}$. In the particular case when $\mathrm{Re}(\tau)= 0$, the two-dimensional convex hull is just a triangle containing the disk of radius $\alpha_{min}$.

From the M-theory perspective, the tip of the cone is associated with a KK tower related to the pure wrapping of the M2-brane on a torus. That is  $n\neq 0$ and $p=q=0$. The other two KK towers in the 2-dimensional convex hull are associated with KK modes of the M2-brane on either of the 1-cycles of the torus. That is, vanishing winding and $p\neq 0$ or $q\neq 0$. In general, for $\mathrm{Re}\tau \neq 0$, the base of the convex hull is associated with states with $p,q\neq 0$ related to $(p,q)$-strings with winding from the type IIB perspective. The string oscillators are located between KK towers in different duality frames, at the boundary of the sphere with radius $\alpha_{min}$. These towers are connected by $\frac{1}{4}$ BPS states at the boundary of the convex hull that have both winding and KK states. These states are located at finite distances in the moduli space. 

Nevertheless, it was proved in \cite{mpgm23} that when considering an M2-brane on $M_9\times T^2$, the KK charges $p,q\neq 0$ are associated with states with $n\neq 0$, being the latter the condition that ensures the discreteness of the spectrum \cite{Boulton}. The relatively prime numbers $p,q$ could be zero and $n\neq 0$, and we will still have an M2-brane with a discrete spectrum. Are the states with $p,q\neq 0$ and vanishing wrapping associated with an M2-brane with discrete spectrum? 

\subsection{The M-theory origin of the towers of massless states}

Let us consider the mass operator of the type IIB $(p,q)$-string on a circle given by the expression \eqref{massoppq}. This mass operator can be obtained via double-dimensional reduction from the M2-brane with nonvanishing wrapping on a torus. 

It can be seen that the tower of states associated with the tip of the cone, the KK tower, is related to the wrapping number of the M2-brane, $n$, from the M-theory perspective, as expected. Indeed, the mass operator of the wrapped QM2-brane with $p=q=0$ given by \eqref{massoppq2}, is associated with a theory with a good quantum description as discreteness of the supersymmetric spectrum.

The states at the base of the convex hull are associated with the winding tower of the ($p,q$)-string from the type IIB perspective. From the M-theory perspective, we have have shown in the previous section that the $T_U$ dual of the QM2-brane  with $p,q\neq 0$ and vanishing wrapping also has a purely discrete spectrum. The integer $n\neq 0$, ensured by the worldvolume flux condition, is associated with the KK term in the dual formulation. The worldvolume theory of this M2-brane is known.

Moreover, the towers associated with the oscillators, which were not computed in \cite{Schwarz6}, were shown in \cite{mpgm23} to be directly related to the membrane excitations of the QM2-brane on a torus.

Consequently, all the towers ensuring the Convex Hull Distance Conjecture in type II maximal supergravity are associated with supersymmetric M2-branes with a discrete supersymmetric spectrum. In other words, the origin in M-theory of the towers is related to well-behaved M2-branes with fluxes, QM2-branes, whose worldvolume and bundle description is known.


\section{Summary and conclusions}
\label{sec5}

We discuss the Convex Hull Distance Conjecture on type II maximal supergravity in nine dimensions from the point of view of the QM2-branes, which corresponds to supersymmetric M2-branes with a discrete spectrum.

From an M-theory perspective, the KK towers appearing at the infinite distance limit in the moduli space were related to the wrapped M2-brane on a torus without KK charges, and to the KK modes of the M2-branes without wrapping. These KK towers were associated with decompactification of one compact dimension.

We find that all these towers that ensure the Swampland Distance Conjecture in this particular case, are directly related to M2-branes with good quantum behaviour, QM2-branes, showing discreteness in the supersymmetric mass operator spectra. The worldvolume flux condition ensures the appearance of these towers at the asymptotic region with the precise decay rate. The local Hamiltonian and the bundle description of these M2-branes are well-known. On one hand, we have the M2-brane with pure central charge and no KK modes. On the other hand, we have the $T_U$ dual M2-brane with pure KK modes and vanishing wrapping number.

Moreover, the states located at the boundary of the convex hull and finite distances in the moduli space are also directly related to M2-branes with discrete spectra. These states are characterized by having both winding and KK modes. As the local and global description of the M2-branes leading to those states is well-known, it would be interesting to check if there is something that we could say about the interior of the moduli space in this simple scenario. This will be studied elsewhere.

In this work, we have focused on the Convex Hull Distance Conjecture. Nevertheless, we consider that the species scale and weak gravity convex hull for type II maximal supergravity can also be related to supersymmetric M2-branes with discrete spectra.

\section*{Acknowledgements}
 MPGM is partially supported by the PID2021-125700NB-C21 and PID2024-155685NB-C21 MCI Spanish Grants and by the University of La Rioja project REGI2025/41. AR thanks to Programa Regional MATH-AMSUD 240048. CLH is supported by FONDECYT postdoctorado No. 3250615 2025. 

\appendix

\end{document}